\documentclass{modegs}                  

\usepackage{times}                   

\usepackage{graphicx}

\printfigures              
\begin{document}
\title{Chaotic advection of reacting substances: 
Plankton dynamics on a meandering jet}

%
%
\author{Crist\'{o}bal L\'{o}pez}
\author[1,2]{Zolt\'an Neufeld} 
\author{Emilio Hern\'{a}ndez-Garc\'{\i}a}
\affil{Instituto Mediterr\'aneo de Estudios Avanzados (IMEDEA), 
CSIC-Universitat de les Illes Balears, E-07071 Palma de Mallorca, 
Spain} 
\author{Peter H. Haynes}
\affil{Department of Applied Mathematics and Theoretical Physics,  
University of Cambridge, Silver Street, Cambridge CB3 9EW, UK} 
%
%
%
%
\date{Manuscript version from 11 June 1999}

\journal{\PCE}       
%
%
\firstauthor{L\'{o}pez} \proofs{Crist\'{o}bal L\'{o}pez
\\IMEDEA, CSIC-Universitat de les Illes Balears\\E-07071 Palma de 
Mallorca, Spain} \offsets{Crist\'{o}bal L\'{o}pez\\IMEDEA, 
CSIC-Universitat de les Illes Balears\\E-07071 Palma de Mallorca, 
Spain} 

\msnumber{} \received{} \accepted{} 

\maketitle

%
%
%

\begin{abstract}
We study the  spatial patterns formed by interacting populations 
or reacting chemicals under the influence of chaotic flows. In 
particular, we have considered a three-component model of plankton 
dynamics advected by a meandering jet. We report general results, 
stressing the existence of a smooth-filamental transition in the 
concentration patterns depending on the relative strength of the 
stirring by the chaotic flow and the relaxation properties of  
planktonic dynamical system. Patterns obtained in open and closed 
flows are compared. 
\end{abstract}

\section{Introduction}
\label{sec:intro}


The transport of biologically or chemically active substances by a 
fluid flow is a problem of great geophysical relevance. Important  
examples arise in the study of atmospheric advection of reactive 
pollutants or chemicals, such as ozone, N$_2$O \citep{ozone}, or 
in the  dynamics of plankton populations in ocean currents 
\citep{plankton}. The inhomogeneous nature of the resulting 
spatial distributions was recognized some time ago \citep[and 
references therein]{patchiness}. More recently, satellite remote 
sensing and detailed numerical simulations identify filaments, 
irregular patches, sharp gradients, and other complex structures 
involving a wide range of spatial scales in the concentration 
patterns. In the case of atmospheric chemistry, the presence of 
strong concentration gradients has been shown to have profound 
impact on global chemical time-scales \citep{legras}. On-site 
measurements and data analysis of the chemical or biological 
fields have confirmed their fractal or multifractal character 
\citep{multifractal1,multifractal2,multifractalO3,tuck}. 

In the case of plankton communities, {\sl patchiness} has been 
variously attributed to the interplay of diffusion and biological 
growth, oceanic turbulence, diffusive instabilities, and nutrient 
or biological inhomogeneities \citep{mackas}. Advection by 
unsteady fluid flow and predator-prey interactions (formally 
equivalent to chemical reaction) are emerging as two key 
ingredients able to reproduce the main qualitative features of 
plankton patchiness \citep{Abraham}. 

The `chaotic advection' paradigm has been shown to be a useful 
approach to understand geophysical transport processes at large 
scales \citep{peter}. Briefly, chaotic advection (or Lagrangian 
chaos)\citep{aref} refers to the Lagrangian complex motion of fluid parcels 
arising from a flow which is not turbulent in the Eulerian 
description. Lagrangian chaotic flows are much simpler than 
turbulent ones, being thus more accessible to analytical 
characterization and understanding. They retain however many of 
the qualitative features relevant to transport and mixing 
processes in complex geophysical flows.  

Though the properties of inert passive tracer fields under chaotic 
advection have been widely studied 
 \citep{Antonsen}  
 much less is known about biologically or 
chemically evolving reactant distributions. 
Nonetheless, some 
results have been recently obtained, as for example in reactions 
of the type $A+B \to C$ in closed chaotic flows \citep{Metcalfe} 
and in open chaotic flows \citep{Toroczkai}. Recently, some of us 
\citep{PRL} considered the general case of stable chemical 
dynamics in closed chaotic flows in the limit of small diffusion 
and in the presence of an external spatially non-homogeneous 
source of one of the chemical components. The main result was that 
the relationship between the rate at which the chemical dynamics 
approaches local equilibrium with the chemical source and the 
characteristic time scale of the stirring by the
 chaotic flow determines the fractal or 
non-fractal character of the long-time distribution. The faster 
the stirring is, the more irregular is the pattern. 

The purpose of this Paper is to apply and verify the general 
results above in a concrete model of plankton dynamics in flows of 
geophysical relevance. In addition we compare structures appearing 
in closed and open flows, stressing the intermittent character of 
irregularities in the open flow case. 
 We expect this result to apply also to other 
situations in atmospheric or oceanic chemistry.  

The paper is organized as follows: next Section summarizes the 
general theoretical results obtained by \citet{PRL}. The 
particular plankton dynamics and the two different flows subject 
of our study are presented in Sect.~\ref{sec:models}. They are 
variations of a kinematic model for a two-dimensional meandering 
jet, leading to a closed and an open flow model. Section 
\ref{sec:closed} describes numerical results for the closed flow 
case, whereas Sect.~\ref{sec:open} considers the open flow.  
Finally, Sect.~\ref{sec:conclusions} contains our conclusions. 

\section{General results}
\label{sec:general}

The temporal evolution of reacting fields is determined by 
advection-reaction-diffusion equations. Advection because they are 
under the influence of a flow, reaction because we consider 
species interacting with themselves and/or with the carrying 
medium. Diffusion because 
turbulent or molecular random motion 
smoothes out the smallest scales. For the case of an  
incompressible velocity field ${\bf v}({\bf r},t)$, the standard 
form of these equations is 
\begin{eqnarray}
\frac{\partial C_i({\bf r},t)}{\partial t} &+& {\bf v} ({\bf 
r},t)\cdot \nabla  C_i({\bf r},t) = \nonumber 
\\& &f_i(C_1,...,C_N,{\bf r})+ \nu \nabla ^2 C_i({\bf r},t) 
\end{eqnarray}
where $C_i({\bf r},t)$, $i=1,...N$, are interacting chemical or 
biological fields advected by the flow ${\bf v}({\bf r},t)$, 
$f_i(C_1,...,C_N,{\bf r})$ are the functions accounting for the 
interaction of the fields (e.g. chemical reactions or 
predator-prey interactions). Diffusion effects are only important 
at small scales and we will neglect them in the following. In this 
limit of zero diffusion $\nu \to 0$  
 the above description can be 
recast in Lagrangian form:  
\begin{equation}
\frac{d{\bf \hat r}}{dt}={\bf v}({\bf \hat r},t)
\label{flow}
\end{equation}
\begin{equation}
\frac{d{C_i}}{dt}={f_i}\left(C_1,C_2,..,C_N,{\bf r}={\bf 
\hat r}(t)\right),\; i=1,.,N, 
\label{chemical}
\end{equation} 
where the second set of equations describes the chemical or 
population dynamics inside a fluid parcel that is being advected 
by the flow according to the first equation. In the absence of 
diffusion, a coupling between the flow 
and the chemical/biological evolution can only appear as a 
consequence of the spatial dependence of the $f_i(C_1,...,C_N,{\bf 
r})$ functions. This spatial dependence describes non-homogeneous 
sources or sinks for the chemical reactants or spatially 
non-homogeneous reaction or reproduction rates. Such 
inhomogeneities may arise naturally from a variety of processes 
such as localized upwelling, inhomogeneous solar irradiation, or 
river run-off, to name a few. 

From now on, the incompressible flow ${\bf v}({\bf r},t)$ will be 
assumed to be two-dimensional and time dependent. This situation 
generally leads  to 
chaotic advection. For simplicity, our general arguments will be 
stated for the case in which ${\bf v}({\bf r},t)$ satisfies the 
technical requirement of {\sl hyperbolicity}, but in the examples
less restrictive flows will be used. The most salient feature of 
advection by a chaotic flow is sensibility to initial conditions, 
that is, fluid particles initially close typically diverge in time 
at a rate given by the maximum Lyapunov exponent of the flow 
$\lambda_F>0$: 
\begin{equation}
|{\bf \delta r}(t)|\sim|{\bf \delta r}(0)| e^{\lambda_Ft} \ .
\label{lyapunov}\end{equation}
Equation (\ref{lyapunov}) is valid for nearly all the initial 
orientations of the initial particle separation ${\bf \delta 
r}(0)$. However, the incompressibility condition implies that 
there is a particular orientation of the initial separations ${\bf 
\delta r_c}(0)$ for which the two trajectories approach each other: $|{\bf 
\delta r}(t)|\sim|{\bf \delta r_c}(0)| e^{\lambda_F't}$, with 
$\lambda_F'=- \lambda_F$.

The general class of chemical reactions studied by \citet{PRL} was 
the one leading to stable local equilibrium in the absence of flow 
or, in terms of the chemical dynamical subsystem (\ref{chemical}), 
the dynamics approaching a unique fixed point for each constant 
position ${\bf r}$. This means that there are not chemical 
instabilities nor chemical chaos, and that the concentrations tend 
to approach a value determined at each point by the sources in 
$f_i$. In the presence of advection by the flow ${\bf r}={\bf 
\hat r}(t)$, the relaxation process is altered, but can be 
characterized by the value of the maximum Lyapunov exponent 
$\lambda_C$ of the chemical subsystem (\ref{chemical}), which we 
assume to remain negative. 

For characterizing the spatial structure of the $C_i$ fields we 
calculate the difference 
\begin{equation}
\delta C_i= C_i({\bf r}+\delta {\bf r},t)-C_i({\bf r},t) \;\;. 
  \end{equation}

\noindent Inserting this expression, for $|{\bf \delta r}|$ small 
enough, into the equation for the chemical dynamics \citep[details 
can be found in][]{PRL} we can obtain the evolution of the 
gradients of the chemical field at long times: 
\begin{eqnarray}
   \nabla C_i({\bf r},t) &\approx& 
   \sum_{j=1}^N     
   \nabla (C_j^0 \cdot {\bf V}_j^0) {\bf V}_i^0
   e^{(\lambda_F+\lambda_C)t}         \nonumber \\
   +  \int_0^t ds &\sum_{j=1}^N& (\nabla f_j({\bf \hat r}(s)) \cdot
    {\bf V}_j^{s}) {\bf V}_i^{s} 
   e^{(\lambda_F+\lambda_C)(t-s)}    \ , 
\label{monster}
   \end{eqnarray}
where the vectors ${\bf V}_i^t \equiv {\bf n}\left({\bf 
\hat r}(t)\right) v_i$ are combinations of the vectors ${\bf n}({\bf r 
})$ pointing at each point along the most contracting direction of 
the flow and of the vector $\{v_i,i=1,...,N\}$ associated to the 
contracting direction in the chemical subspace. ${\bf \hat r}(s)$ is the 
trajectory ending at ${\bf r}$ at time $t$. It is important to 
realize that Eq.~(\ref{monster}) gives the long time behavior of 
$\nabla C_i$ correctly in all but in one direction. The 
directional derivative of $ C_i$ in the most expanding direction 
should be obtained with (\ref{monster}) but replacing $\lambda_F$ 
by $\lambda_F'=-\lambda_F$, and the vectors ${\bf n} $ by the ones  
associated to the expanding direction.  

  The convergence of the gradients for $t\to \infty$ depends on the
sign of the exponent $\lambda_F + \lambda_C$. There are two 
possibilities: 
\begin{itemize}
\item  If $\lambda_F + \lambda_C < 0 $ then the convergence of the
chemical dynamics towards local equilibrium is stronger than the 
effect of the chaotic flow on the fluid particles. Gradients are 
finite so that a smooth asymptotic distribution is attained by the  
chemical fields. 
\item  If $\lambda_F + \lambda_C > 0 $ then in the $t\to \infty$ limit 
the chemical pattern becomes {\it nowhere differentiable}. An 
irregular structure with fractal properties is 
developed. Remember however that at each point there is a 
direction for which  $\lambda_F + \lambda_C$ should be substituted 
by $-\lambda_F + \lambda_C$, always negative, in (\ref{monster}). 
In this direction derivatives are finite and the field is smooth. 

Thus the resulting structure is {\sl filamental}, i.e.,
irregular in all directions except in one along which it is smooth.
This one corresponds to the direction of the filaments lying along
the unstable foliation of the chaotic advection.
 The fractal 
characteristics of the filamental structure in the closed flow 
case were investigated by \citet{PRL}. We will see however that 
there are differences between the closed and the open case, which 
will be discussed in Sects.~\ref{sec:closed} and \ref{sec:open}. 
\end{itemize}

\section{The plankton and the jet models}
\label{sec:models}

In the numerical investigations below we will consider a simple 
model of plankton dynamics immersed in a meandering jet flow.  

This plankton model, used by \citet{Abraham} and related to the 
one used by \citet{turing}, considers explicitly three trophic 
levels: the nutrient content of a water parcel, described in terms 
of its carrying capacity $C$ (defined as the maximum phytoplankton 
content it can support in the absence of grazing), the 
phytoplankton biomass $P$, and the zooplankton $Z$. The Lagrangian 
`chemical'  subsystem (\ref{chemical}) reads: 
\begin{eqnarray}
\frac{dC}{dt}&=& \alpha \left(   C_0({\bf r})-C \right) 
\label{Ct}\\ 
\frac{dP}{dt}&=&P\left( 1-\frac{P}{C}\right)-PZ   
\label{Pt}\\ 
\frac{dZ}{dt}&=&PZ-\delta Z^2 \label{Zt}\ . 
\end{eqnarray}
All terms have been adimensionalized to keep a minimal number of 
parameters. Equation (\ref{Ct}) describes the relaxation of the 
carrying capacity, at a rate $\alpha$, towards an inhomogeneous 
shape $C_0({\bf r})$. This will be the only explicitly 
inhomogeneous term in the model, and describes a spatially 
dependent nutrient input, arising from some topography-determined 
upwelling distribution or latitude dependent illumination, for 
example. The first terms in Eq.~(\ref{Pt}) describe phytoplankton 
logistic growth, whereas the last one models predation by 
zooplankton. This effect gives also rise to the first term in 
(\ref{Zt}). The term containing $\delta$, the zooplankton 
mortality, describes zooplankton death produced by higher trophic 
levels. The only stable fixed point of model (\ref{Ct})-(\ref{Zt}) 
is given by $C^*=C_0({\bf r})$, $P^*= {C_0 \delta }/({\delta + 
C_0})$, and $Z^*={P^*}/{\delta}$. 

The model flow will be given by the following streamfunction 
\citep{jet}: 
\begin{eqnarray}
&&\phi(x,y)   =   \nonumber \\ &&1 - \tanh \left(\frac{y-B(t) 
\cos\left[ k(x-ct)\right]}{\left( 1+k^2 B(t)^2 \sin^2 
\left[k(x-ct)\right] \right)^{\frac{1}{2}}} \right) 
\label{jetstreamfunction}
\end{eqnarray}
It describes a jet flowing eastwards, with meanders in the 
North-South direction which are themselves advected by the jet at 
a phase velocity $c$. $B(t)$ and $k$ are the (properly 
adimensionalized) amplitude and wavenumber of the undulation in 
the streamfunction. 

The motion of the tracer particles (the dynamical system 
(\ref{flow}) ) is given by  
\begin{eqnarray}
\frac{dx}{dt}&=&-\frac{\partial \phi}{\partial y} \nonumber\\ 
\frac{dy}{dt}&=&\frac{\partial \phi}{\partial x} 
\label{streamfunction}
\end{eqnarray}

If the amplitude $B$ of the meanders is time-independent, a simple 
change in the frame of reference renders the flow time-independent 
and Eq.~(\ref{streamfunction}) defines a non-chaotic integrable 
dynamical system. Chaotic advection appears in this model if $B$ 
is made to vary in time, for example periodically: 
\begin{equation}
B(t)=B_0+\epsilon \cos(\omega t+\theta) \ .
\label{Bt}
\end{equation}
Following \citet{vulpiani}, we use the parameter values $B_0=1.2, 
c=0.12, k={2\pi}/{L_x}, L_x=7.5, \omega=0.4,\epsilon=0.3 $ and $\theta 
=\frac{\pi}{2}$.
 These values guarantee the existence of `large 
scale chaos', i.e, the possibility that a test particle crosses 
the jet passing from North to South or viceversa. This is weaker 
than the requirement of hyperbolicity, but is enough to illustrate 
the general aspects of our theory.  

The natural interpretation of the jet-flow just introduced is as 
an open flow: it advects most of the fluid particles from 
$x=-\infty$ towards $x=\infty$.
We will localize the source term near the origin of coordinates:
\begin{eqnarray}
C_0(x,y) = \left\{ \begin{array}{l}
{1+A\sin(2\pi x/L_x)\sin(2\pi y/L_y)}\\
{\;\;\;\;\;\;{\rm if} -L_x \leq x \leq L_x}\\
{}\\
{{0}\;\;\; \rm{elsewhere}}
\end{array} \right.
\end{eqnarray}

 In this way, there is inhomogeneous 
nutrient input just near the origin, and capacity and plankton 
concentration in the fluid particles will decay as they are 
advected downstream.  

A quite different class of natural flows are closed ones, e.g.  
recirculating flows in closed basins. Our jet model flow can be 
made closed simply by imposing periodic boundary conditions at the 
ends of the interval $-L_x<x<L_x$. Particles leaving  
 the region through the right boundary are 
reinjected from the left. In this way nutrients are injected and 
extracted continuously from fluid elements as they traverse the 
different regions of the source
\begin{equation}
C_0(x,y)=1+A \sin(2\pi x/L_x)\sin(2\pi 
y/L_y)
\end{equation}
 This was the situation considered by \citet{PRL}. We will 
see that different structures develop in  the open and in the 
closed situation. 

\section{Closed flows}
\label{sec:closed}

Numerically we proceed by integrating backwards in time Eq. (\ref{flow}) 
 with initial coordinates on a rectangular grid ($300 \times 150 $)
and then the chemical field for each point is obtained by 
integrating (\ref{chemical}) forward in time along the fluid 
trajectories so generated. 

\begin{figure}
\figbox*{}{}{
\includegraphics[width=.8\columnwidth]{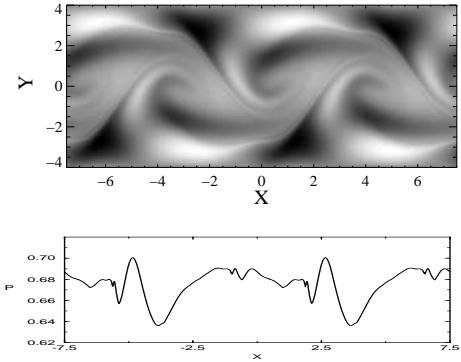}
} \caption{Phytoplankton smooth pattern in the closed flow, with an horizontal  
transect taken along $y=0.8$} 
\label{figure1}
\end{figure}

\begin{figure}
\figbox*{}{}{ 
\includegraphics[width=.8\columnwidth]{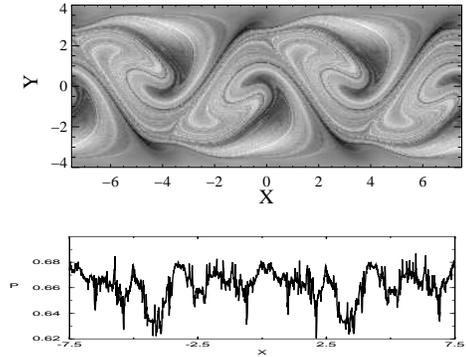}
} \caption{Phytoplankton filamental pattern in the closed flow, with a transect 
along $y=0.8$ } 
\label{figure2}
\end{figure}

Figure \ref{figure1} shows a snapshot of the long-time 
phytoplankton distribution in the closed flow case for parameter 
values $\alpha = 0.25, \delta=2.0, L_y=4. $ and $A=0.2$. In this 
case $\lambda_F+\lambda_C<0$, therefore the distribution is 
smooth. A transect along the line $y=0.8$ is also shown. Taking 
$\alpha = 0.025$, so that now $\lambda_F+\lambda_C>0$ we obtain 
the distribution in Fig. \ref{figure2}. A complex filamental 
structure is clearly seen, in agreement with our theoretical 
arguments. The fractal nature of the pattern is also seen in the 
horizontal cut presented also in Fig.~\ref{figure2}. A H\"{o}lder 
exponent of $|\lambda_C | / \lambda_F$ was predicted for these 
kind of transects in \citep{PRL}. This implies plankton-variance 
power spectrum decaying as $k^{-\beta}$, with $\beta=1+2|\lambda_C 
| / \lambda_F$. Thus $\beta $ is in the range  $]1,3]$ which
agrees with field observations of plankton distributions \citep{multifractal2}.

\section{Open flows}
\label{sec:open}

Contrarily to closed flows, open flows are characterized by
unbounded trajectories of fluid particles.  Typical fluid 
particles enter and, after some time, leave the region of the 
nutrient source ($C_0({\bf r})>0$). Thus, most  fluid elements 
have only spent the most recent part of their trajectories inside 
the active region, with the rest of their evolution spent in 
regions with no spatial dependence of the nutrient input. 
 For this part of the trajectory
the gradient $\nabla f_j$ in Eq.~(\ref{monster}) vanishes, implying that
there would not be  divergence of
the gradients in the long-time limit. 

It is well known from the study of chaotic advection in open flows 
\citep{openflows}, that while most of the particles spend only a 
finite amount of time in selected bounded regions  of the flow, 
typically these regions contain also bounded orbits in which some 
of the particles can stay forever. Although the chaotic set formed 
by the bounded orbits is a fractal set of measure zero, particles 
visiting the vicinity of the stable manifold of this set can still 
spend arbitrarily long time in the selected bounded region. This 
leads to the formation of characteristic fractal patterns in the 
advection dynamics even in the case of passive particles, as it 
was shown in numerical studies \citep{jung} and laboratory 
experiments of open flows \citep{experim}. 

\begin{figure}
\figbox*{}{}{ 
\includegraphics[width=.8\columnwidth]{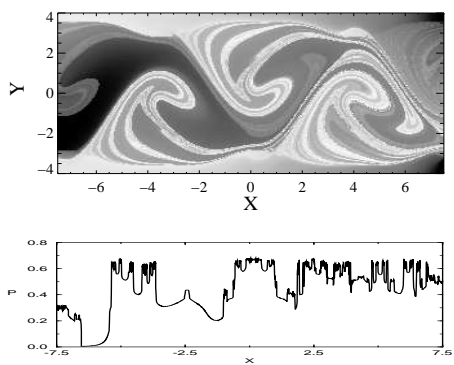}
} \caption{Phytoplankton filamental pattern in the open flow, with a transect 
along $y=0.8$. } 
\label{figure3}
\end{figure}

For the particles that have spent infinitely long time
in the source region, their chemical/biological evolution is equivalent
to the one in a closed flow  with the possibility of diverging 
gradients. The only difference is that now the values of $\lambda_F$
and $\lambda_C$ are those corresponding to 
the chaotic set of bounded 
orbits that never leave the biologically active region.
The smooth-filamental transition observed in the previous
Section for the closed flow will occur here only on this
fractal set, which will always be surrounded by
a smooth distribution. This results in a strongly intermittent
character of the filamental field.
Figure \ref{figure3} shows a phytoplankton pattern for
parameter values 
$\alpha = 0.025, \delta=2.0 $ and $A=0.2$,
 and a transect crossing it, in the open-flow case. The
inhomogeneity in the filamental structure is obvious, with 
singularities in a set recognized as the unstable manifold of the 
chaotic set formed by bounded orbits in the source region. Smooth 
structures are also obtained when $\lambda_F+\lambda_C<0$. The 
proper description of the resulting structures should use the 
concept of multifractality, that is,  inhomogeneous distribution 
of fractal properties. In fact, even in the closed flow case, at 
finite times the flow Lyapunov exponent $\lambda_F$ would have 
space-dependent finite-time corrections, that will need to be 
taken into account in a proper generalization of (\ref{monster}).  
A quantitative description of these multifractal filamental 
structures will be presented elsewhere \citep{PRE}.

\section{Conclusions}
\label{sec:conclusions}

The  spatial patterns formed by interacting populations under the 
influence of chaotic flows have been studied. In particular, we 
have considered a coupled model of `nutrient', phytoplankton and 
zooplankton concentrations advected by two (open and closed) 
jet-like flows. General results have been reported for arbitrary 
chaotic flows, stressing the existence of a smooth-filamental 
transition depending on the relative strength of the maximum 
Lyapunov exponent of the flow and the one corresponding to the 
planktonic dynamical system. Patterns obtained for open and closed 
chaotic flows are different because of the transient or permanent 
character of the biological activity. 
 Comparison of the 
structures for the different fields (capacity, phytoplankton and 
zooplankton), and quantitative description of their multifractal 
properties will be presented elsewhere \citep{PRE}.  

The models considered here are extreme simplifications of real 
biological and geophysical situations. We expect however that the 
main qualitative features found here, namely the possibility of 
finding smooth or filamental patterns depending on stirring and 
relaxation rates, and the increased inhomogeneities in open flows, 
to be present in more realistic chemical or biological transport 
situations.  

\balance 

\begin{acknowledgements}
Helpful discussions with Tam\'as T\'el are acknowledged. This work 
was supported by  CICYT project MAR98-0840. Z.N. was supported by 
an European Science Foundation/TAO ({\it Transport Processes in the 
Atmosphere and the Oceans }) Exchange Grant. 

\end{acknowledgements}


\end{document}